\documentclass[journal]{IEEEtran}
\usepackage{graphicx}
\usepackage{epstopdf}
\usepackage{amssymb}

\usepackage{color}
\graphicspath{{figures/}}
\usepackage{tabularx}
\usepackage{multirow}
\usepackage{booktabs}
\usepackage{enumitem}
\usepackage{flushend}
\usepackage{amsmath}

\begin{document}

\title{Multi-materials beam hardening artifacts correction for computed tomography (CT) based on X-ray spectrum estimation}
\author{Wei~Zhao,~
        Dengwang~Li,~
        Kai~Niu,~
        Wenjian~Qin,~
        Hao~Peng$^*$,~
        and~Tianye~Niu$^*$~
\thanks{This work is supported by the Zhejiang Provincial Natural Science Foundation of China (Grant No. LR16F010001), National High-tech R\&D Program for Young Scientists by the Ministry of Science and Technology of China (Grant No. 2015AA020917), National Key Research Plan by the Ministry of Science and Technology of China (Grant No. 2016YFC0104507), Natural Science Foundation of China (NSFC Grant No. 81201091, 61601190, 51305257). \emph{Asterisk indicates corresponding authors.}}
\thanks{W. Zhao is with the Department of Biomedical Engineering, Huazhong University of Science and Technology, Wuhan, 430074 China and with the Department of Radiation Oncology, Stanford University, Stanford, CA, 94305 USA and also with the Key Laboratory of Computer Network and Information Integration, Southeast University, Ministry of Education, Nanjing 210096 China.}
\thanks{D. Li is with the Shandong Province Key Laboratory of Medical Physics and Image Processing Technology, School of Physics and Electronics, Shandong Normal University, Jinan, 250014 China.}
\thanks{K. Niu is with the Department of Medical Physics, University of Wisconsin-Madison, Madison, WI, 53705 USA.}
\thanks{W. Qin is with the Shenzhen Institutes of Advanced Technology, Chinese Academy of Sciences, Shenzhen, 518055 China.}
\thanks{H. Peng is with the Department of Radiation Oncology, Stanford University, Stanford, CA, 94305 USA and also with the Department of Physics, Wuhan University, Wuhan, Hubei, China (e-mail: haopeng@stanford.edu).}
\thanks{T. Niu is with the Sir Run Run Shaw Hospital and Institute of Translational Medicine, Zhejiang University, Hangzhou, 310016 China (e-mail: tyniu@zju.edu.cn).}
}

%



\maketitle

\begin{abstract}

Due to the energy dependent nature of the attenuation coefficient and the polychromaticity of X-ray source, beam hardening effect occurs when X-ray photons penetrate through an object, causing a nonlinear projection data. When a linear reconstruction algorithm, such as filtered backprojection, is applied to reconstruct the projection data, beam hardening artifacts which show as cupping and streaks, are present in the CT image. The aim of this study was to develop a fast and accurate beam hardening correction method which can deal with beam hardening artifacts induced by multi-materials objects. Based on spectrum estimation, the nonlinear attenuation process of the X-ray projection was modeled by reprojecting a template image with the estimated polychromatic spectrum. The template images were obtained by segmenting the uncorrected into different components using a simple segmentation algorithm. By adding the scaled difference of the monochromatic reprojection data and the polychromatic reprojection to the raw projection data, the raw projection data was mapped into the corresponding monochromatic projection data, which was used to reconstruct the beam hardening artifacts corrected images. The algorithm can also be implemented in image-domain which takes the uncorrected image volume as input. In this case, the scaled mapping term was reconstructed to yield a set of artifacts images which can be added directly to the uncorrected images. Numerical simulations, experimental phantom data and animal data which were acquired on a modern diagnostic CT scanner (Discovery CT750 HD, GE Healthcare, WI, USA) and a modern C-Arm CT scanner (Artis Zee, Siemens Healthcare, Forchheim, Germany), respectively, were used to evaluate the proposed method. The results show the proposed method significantly reduced both cupping and streak artifacts, and successfully recovered the Housfield Units (HU) accuracy.

\end{abstract}



%
\IEEEpeerreviewmaketitle

\section{Introduction}

\IEEEPARstart{C}{omputed} tomography (CT) imaging is based on attenuation property of the scanned object. The attenuation property which can be characterized by a linear attenuation coefficient is a function of X-ray photon energy. Typically, as X-ray photon energy increases, the linear attenuation coefficient goes down. Meanwhile, X-ray photons used for diagnosis are usually polychromatic, thus X-ray photons with lower energy are more likely to be absorbed when polychromatic X-ray photons penetrate an object and the fraction of high energy photons goes up, causing an increased mean energy which is called as beam hardening. Based on this nonlinear nature, when a linear reconstruction algorithm, such as filtered backprojection (FBP), is applied to reconstruct the projection data, beam hardening artifacts which show as cupping and streaks, are present in the reconstructed CT image. These artifacts lead to worse visual, contrast reduced and inaccurate CT images~\cite{maki1999}. Images after beam hardening correction (BHC) can show better visualization, yield more accurate segmentation and also provide more accurate quantitative analysis~\cite{kitagawa2010,van2004,wei2011,zhao2016}.

\begin{figure}[t]
    \centering
    \includegraphics[width=3.0 in]{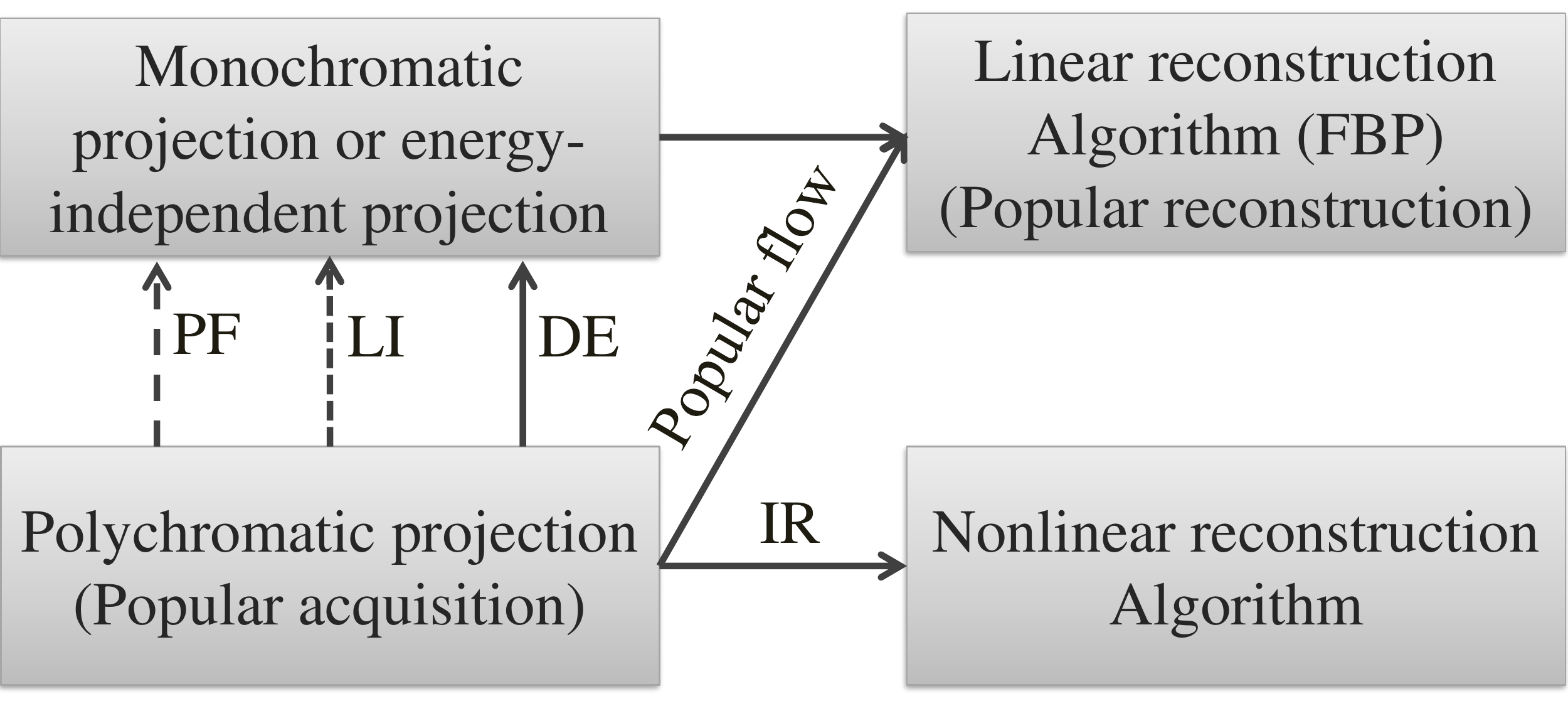}
    \caption{Schemes of beam hardening correction for x-ray CT scanner. The popular work flow for CT is that the nonlinear polychromatic projections acquired by modern CT scanners are reconstructed using a linear filtered backprojection algorithm, such as filtered backprojection (FBP). The inconsistency yields beam hardening artifacts in CT images. To correct the beam hardening artifacts, a possible solution is to map the polychromatic projection into corresponding monochromatic projections or energy-independent projections, which can be implemented using prefiltering (PF), linearization (LI), and dual-energy imaging(DE). Then a linear reconstruction algorithm is applied sequentially to yield beam hardening artifacts reduced image. Alternatively, nonlinear iteration reconstruction (IR) algorithm which takes the polychromatic X-ray spectrum into account can also be used to reconstruct beam hardening reduced image. }
    \vspace{-1em}
    \label{fig:bhc}
\end{figure}

The general schemes of BHC for X-ray CT scanner are shown in Fig.~\ref{fig:bhc}. While most of the modern CT scanners acquire nonlinear polychromatic projections, most of the image reconstruction methods employ FBP algorithm which is a linear reconstruction algorithm. Thus the data acquisition model is not consistent with the reconstruction algorithm, yielding the beam hardening artifacts contaminated CT image. To correct the beam hardening artifacts, a possible solution is to map the polychromatic projection into energy-independent projection or corresponding monochromatic projection. This can be implemented using prefiltering (PF), linearization (LI)~\cite{ritschl2010}, dual-energy decomposition (DE)~\cite{Alvarez1976}. Alternatively, nonlinear iteration reconstruction (IR) algorithm can also yield beam hardening artifacts reduced image by taking the polychromatic attenuation into account~\cite{yan2000,DeMan2001,Elbakri2002}.

Prefiltering is usually just simply adding an additional filter (such as aluminum, copper or filter with K-edge in the X-ray imaging energy range) between the X-ray source and the object to generate a hardened spectrum. The filtered spectrum usually has higher mean energy and narrower energy range than the incident spectrum. It yields beam hardening artifacts mitigated images, however, because of its insufficient correction, this technique is usually employed with other BHC techniques together. Dual-energy method can provide CT images at different monochromatic photon energies from dual kVp scans and these images are beam hardening artifacts free in principle~\cite{Alvarez1976,Stonestrom1981,coleman1985,kalender1986,lin2011,Yu2012}, but due to the extra calibration measurement and CNR reduction at some energy level, dual-energy can only be applied in some specific applications right now~\cite{Yu2012}. BHC methods based on iterative reconstruction try to incorporate the polychromatic attenuation nature into the update procedure to reconstruct beam hardening artifacts free images~\cite{yan2000,DeMan2001,Elbakri2002,elbakri2003,menvielle2005,Brabant2012,cai2013,wu2014,yang2014,jin2015,luo2017,xu2017}. Instead of using a simple monochromatic projection acquisition model,~\cite{DeMan2001} employed a polychromatic acquisition model and decomposed the energy-dependent linear attenuation coefficient into a photonelectric component and a Compton scatter component.~\cite{Elbakri2002} proposed an polychromatic model-based iterative algorithm to estimate the energy-independent density in each voxel. By incorporating the effect of beam hardening into the forward projection of SART update,~\cite{Brabant2012} reported an algorithm which can reconstruct the image and reduce the beam hardening artifacts at the same time. Linearization methods try to map the uncorrected polychromatic projection data into corresponding monochromatic projection data~\cite{Herman1979,Hammersberg1998,VandeCasteele2002,de2004,Gao2006,li2007,Krumm2008,ritschl2010,wei2011,VanGompel2011}. Usually, a calibration measurement is needed to perform the correction. A few linearization methods can correct high-order beam hardening artifacts without calibration measurement~\cite{Krumm2008,Kyriakou2010}.

It has to note that besides the aforementioned BHC methods, to yield beam hardening artifacts free CT image, one may use a photon-counting or an energy-resolved detector to acquire the projection data instead of using an energy-integrating (EI) detector. Since most of the CT scanners employ EI detectors, we will focus on the EI detector in this study.

The interests in correction of beam hardening artifacts induced by multi-material objects (especially bone) began since early days of CT imaging~\cite{joseph1978,Hsieh2000,Krumm2008,so2009,Kyriakou2010,zhang2010,tang2011,VanGompel2011,Brabant2012,Yang2013,lin2014,chen2014,zhaoyunsong2015,luo2017}, however, high order beam hardening artifacts are still present in CT images acquired on modern CT scanners from time to time. In this study, with the incorporation of an accurate X-ray imaging physical model, we developed a fast BHC strategy which can deal with the high order beam hardening artifacts without iterative projection and backprojection steps. Numerical simulations, experimental phantom data and animal data were used to evaluate the proposed method.

\section{Material and methods}
\begin{figure*}[t]
    \centering
    \includegraphics[width=0.8\textwidth]{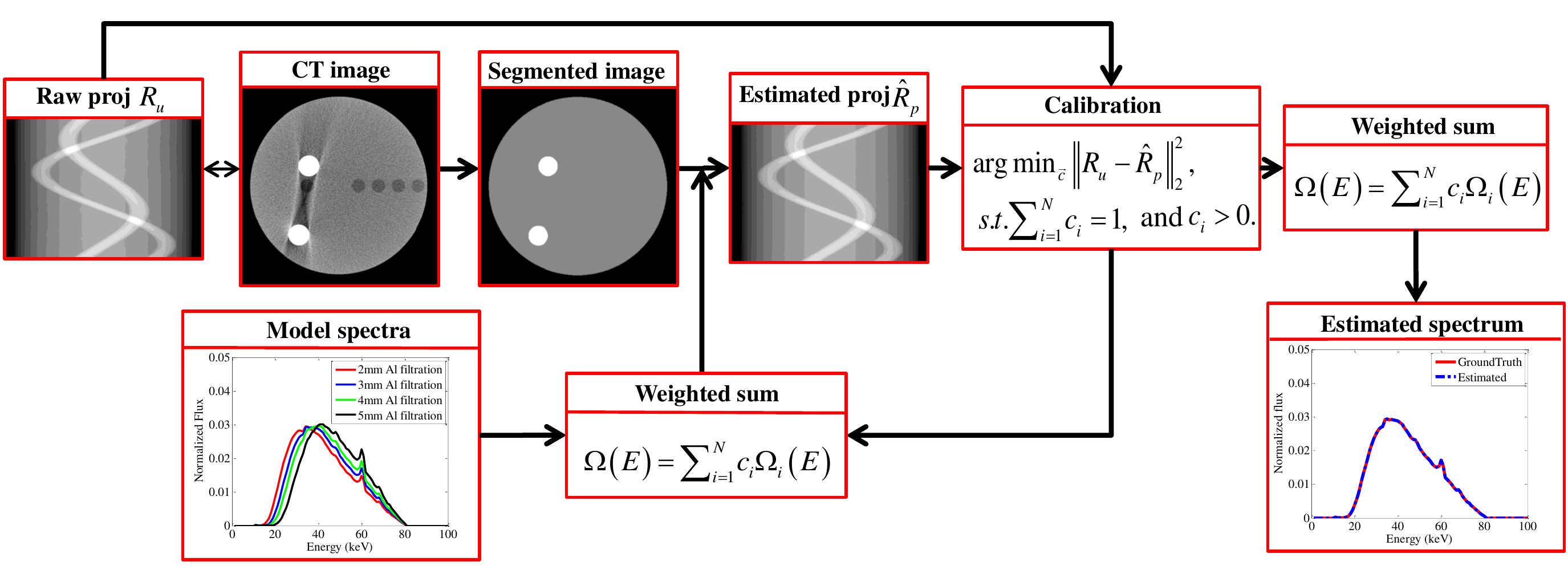}
    \caption{Flowchart of the indirect transmission measurement-based energy spectrum estimation method. The method takes either raw projection data or an image volume as input. If the image volume is the input, the raw projection data can be obtained by a forward projection of the volume. If the raw projection data is the input, a volume is reconstructed. The image volume is then segmented into different components, after which a polychromatic reprojection is performed using the segmented images and a weighted summation of a set of model spectra. The unknown weights are iteratively updated to minimize the difference of the estimated polychromatic reprojection data and the raw projection data. The final spectrum is calculated using the calibrated weights and the model spectra.}
    \vspace{-1em}
    \label{fig:spek}
\end{figure*}
\subsection{Description of algorithm}

When X-ray photons with monochromatic energy $E_{\mathrm{m}}$ are used to scan an object, the detector readout $I_{\mathrm{m}}$ which satisfies the Beer's law can be expressed as
\begin{equation}
I_{\mathrm{m}}=N\, \eta(E_{\mathrm{m}})\,\mathrm{exp}\left[-\int_{0}^{l}\mu(E_{\mathrm{m}},\vec{x})\mathrm{d}\vec{x}\right],
\end{equation}
with $N$ the total number of the incident photons, $\vec{x}$ the spatial position variable, $\mu(E_{\mathrm{m}},\vec{x})$ the linear attenuation coefficient at the energy level $E_{\mathrm{m}}$, $l$ the propagation path length of the photons and $\eta(E_{\mathrm{m}})$ is the energy dependent response of the detector and it can be considered as proportional to photon energy $E_{\mathrm{m}}$ because most CT scanners use EI detectors. With these notations, the flood field $I^0_{\mathrm{m}}$ of the monochromatic X-ray source which is the average number of photons received by the detector with the absent of the object can be computed as
\begin{equation}
I_{\mathrm{m}}^{\mathrm{0}}=N\, \eta(E_{\mathrm{m}}).
\end{equation}
However, in reality, the X-ray photons emitted from an X-ray tube of a CT scanner are polychromatic. In this case, the detector readout $I_{\mathrm{p}}$ should be reformulated into
\begin{equation}\label{equ:polyproj}
I_{\mathrm{p}}
=N\int_{0}^{E_{\mathrm{max}}}\mathrm{d}E\,\Omega(E) \, \eta(E)\,\mathrm{exp}\left[-\int_{0}^{l}\mu(E,\vec{x})\mathrm{d}\vec{x}\right],
\end{equation}
where $\Omega(E)$ is the polychromatic X-ray spectrum and $E_\mathrm{max}$ is the maximum photon energy of the spectrum. The corresponding flood field of the polychromatic X-ray source is
\begin{equation}\label{equ:polyair}
I_{\mathrm{p}}^{\mathrm{0}}=N\int_{0}^{E_{\mathrm{max}}}\mathrm{d}E\,\Omega(E) \, \eta(E).
\end{equation}

Thus the projection in logarithmic domain $R_{\mathrm{m}}$, $R_{\mathrm{p}}$ for the monochromatic and polychromatic photons respectively are then calculated as follows:
\begin{equation}
R_{\mathrm{m}} = -\mathrm{ln}(\frac{I_{\mathrm{m}}}{I_{\mathrm{m}}^{\mathrm{0}}}),
\end{equation}
\begin{equation}\label{equ:polyRaysum}
R_{\mathrm{p}} = -\mathrm{ln}(\frac{I_{\mathrm{p}}}{I_{\mathrm{p}}^{\mathrm{0}}}).
\end{equation}
Note that $R_{\mathrm{m}}, R_{\mathrm{p}}$ are detector pixel dependent and we have dropped the detector index for convenience. The aforementioned mapping-type BHC method (linearization) tries to estimate the difference of the equivalent monochromatic projection $R_{\mathrm{m}}$ and the polychromatic projection $R_{\mathrm{p}}$, i.e.,
\begin{equation}\label{equ:delta}
\begin{split}
\delta_{\mathrm{BHC}} &= R_{\mathrm{m}}-R_{\mathrm{p}}\\
&= \mathrm{log}(\frac{\sum_i \Omega_m (E_i)\eta(E_i)\mathrm{exp}\left[-\sum_j\mu_j(E_i)L_j\right]}{\sum_i \Omega_m (E_i)\eta(E_i)})\\
&-\mathrm{log}(\frac{\sum_i \Omega_p (E_i)\eta(E_i)\mathrm{exp}\left[-\sum_j\mu_j(E_i)L_j\right]}{\sum_i \Omega_p (E_i)\eta(E_i)}).
\end{split}
\end{equation}
Here $i, j$ are the energy and material indexes, respectively. $\Omega_m (E_i), \Omega_p (E_i)$ and $\eta(E_i)$ are the components of monochromatic energy spectrum $\Omega_m$, polychromatic energy spectrum $\Omega_p$ and detector response $\eta$ at the energy level $E_i$, respectively. $\mu_j(E_i)$ represents the attenuation coefficient of the $j$th material at the energy level $E_i$. $L_j$ is the propagation path length in the $j$th material and it can be calculated using a ray-tracing algorithm. For BHC, $\delta_{\mathrm{BHC}}$ can be calculated by polychromatic reprojecting (i.e., begin with an image and produce projections) a template that has similar attenuation properties as the scanned object and then added directly to the uncorrected projection data $R_{\mathrm{u}}$ which is acquired by a detector. This completes the mapping process from polychromatic to monochromatic to yield the beam hardening corrected projection data $R_{\mathrm{c}}$, i.e.
\begin{equation}\label{equ:corretionFormula}
R_{\mathrm{c}}=R_{\mathrm{u}}+\delta_{\mathrm{BHC}}.
\end{equation}
Different from $R_{\mathrm{p}}$ and $R_{\mathrm{m}}$ which are numerically calculated using polychromatic and monochromatic photons, respectively, $R_{\mathrm{u}}$ is a realistic measurement and should include primary and scatter signals, as well as noise. In the proposed algorithm, the difference between the polychromatic reprojection $R_{\mathrm{p}}$ and the monochromatic reprojection $R_{\mathrm{m}}$ of the template which was segmented from the uncorrected CT images, was used to characterize the true mapping term $\delta_{\mathrm{BHC}}$ of the object. Then the performance of the algorithm was dependent on the accuracy of the spectrum that was used to perform the polychromatic reprojection. 
The energy level of the monochromatic spectrum which was usually selected to be close to the effective energy of the estimated polychromatic spectrum and the estimation of the polychromatic spectrum will be presented in the next subsection.

In realistic applications, it is quite often that one does not have access to the raw projection data on real scanners, especially for diagnostic CT scanners. In this case, beam hardening correction can not be performed in projection domain. However, based on the linear property of Radon transform $\Re$, the reconstructed CT images can be expressed as
\begin{equation}\label{equ:corretionFormula2}
\begin{split}
f_{\mathrm{c}}&=\Re^{-1}(R_{\mathrm{c}})=\Re^{-1}(R_{\mathrm{u}}+\delta_{\mathrm{BHC}})\\
&=\Re^{-1}(R_{\mathrm{u}})+\Re^{-1}(\delta_{\mathrm{BHC}}) = f_{\mathrm{u}} + \Delta f_{\mathrm{BHC}}.
\end{split}
\end{equation}
Based on~(\ref{equ:corretionFormula2}), to fully address the artifacts in image domain, $\delta_{\mathrm{BHC}}$ needs to be accurately calculated, suggesting an adequate model of the spectrum.

In this study, the segmentation procedure was performed using the OTSU method~\cite{Otsu1975}. For one level segmentation, the method assumed that the image contains two classes of pixels, it then calculated the optimum threshold separating the two classes such that their intra-class variance was minimal, or equivalently, such that their inter-class variance was maximal. In the numerical implementation, we searched for the threshold $t$ that minimizes the intra-class variance, which was defined as a weighted summation of variance of the two classes:
\begin{equation}\label{equ:temp}
 \sigma_w^2(t)= w_1(t)\sigma_1^2(t) + w_2(t)\sigma_2^2(t).
\end{equation}
Here the weights $w_{i=\{1,2\}}$ were the cumulative probabilities of the classes and $\sigma_{i=\{1,2\}}$ were their variances. 

\subsection{Polychromatic spectrum estimation}

\begin{figure}[]
    \centering
    \includegraphics[width=3.0 in]{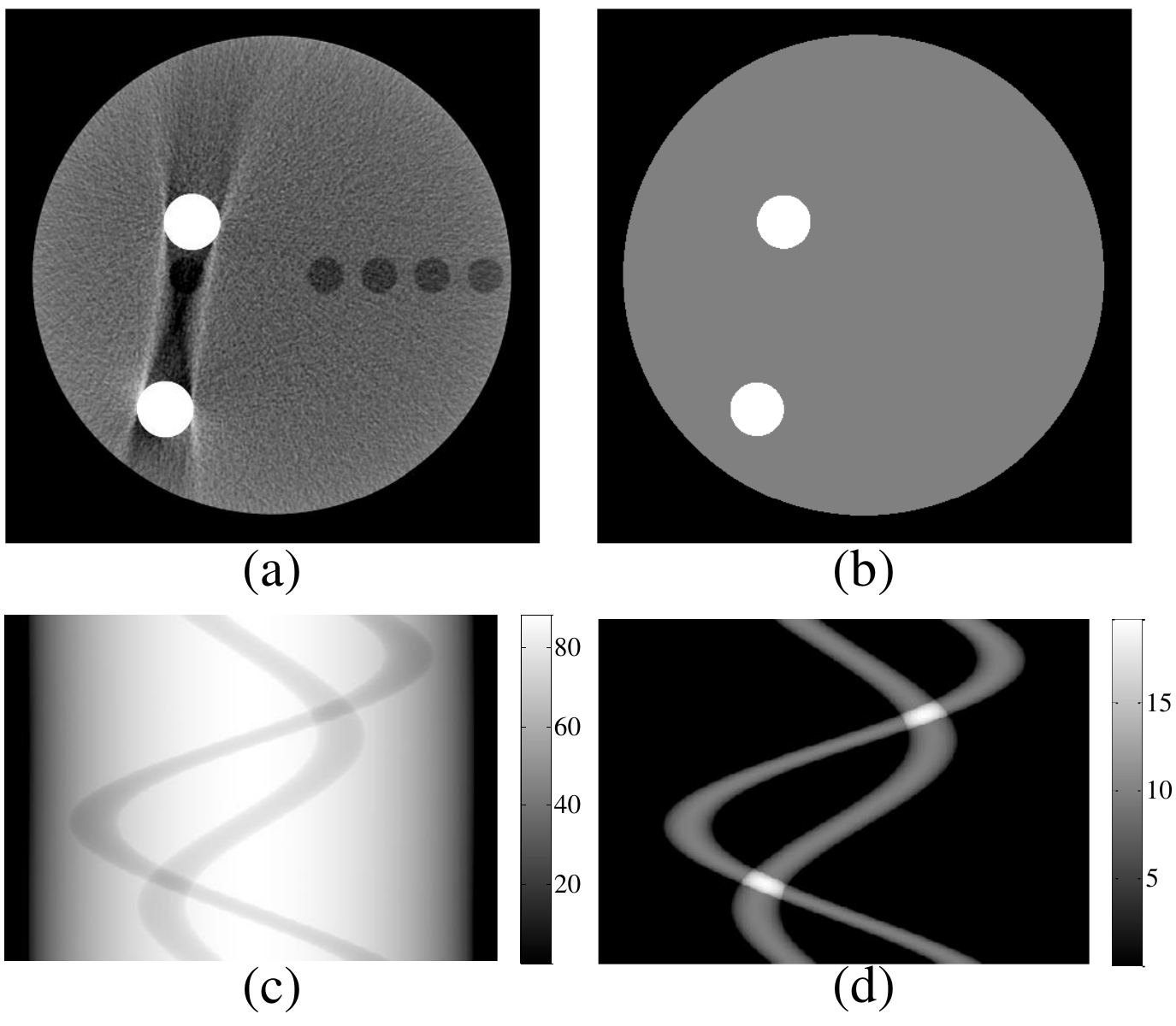}
    \caption{Uncorrected image, segmented image and corresponding propagation path length in the segmented materials. (a) Uncorrected image, (b) two components (aluminum and PMMA) segmentation image, (c) propagation path length in PMMA, (d) propagation path length in aluminum. Note that the water inserts can not be distinguished from the PMMA background. The unit for propagation path lengths is millimeter. }
    \label{fig:reproj}
\end{figure}

To accurately estimate $\delta_{\mathrm{BHC}}$ which was the difference of the monochromatic reprojection data and the polychromatic reprojection data, the polychromatic spectrum should be modeled precisely. In this study, an indirect transmission measurement-based spectrum estimation method was employed to estimate an effective spectrum which can model the polychromatic attenuation process of the projection data~\cite{zhao2015}. The flowchart of the spectrum estimation method was illustrated in Fig.~\ref{fig:spek}.

The method started with either the raw projection data or a beam hardening artifacts contaminated volume. If the image volume was the input, the raw projection data can be obtained by a forward projection using the volume. If the raw projection data was the input, a volume was reconstructed using the raw projection data. The first step of the spectrum estimation was to segment the uncorrected images into different components. By calculating the propagation path length (PPL) for each of the segmented components for each detector pixel using a GPU-based ray-tracing algorithm (as illustrated in Fig.~\ref{fig:reproj}), we can generate a set of polychromatic reprojection data $\hat{R}_{\mathrm{p}}$ using~(\ref{equ:polyproj}) and~(\ref{equ:polyRaysum}) with the PPLs and an estimated polychromatic spectrum. The estimated spectrum was then iteratively updated to minimize the difference of the measured raw projection data $R_{\mathrm{u}}$ and the polychromatic reprojection data $\hat{R}_{\mathrm{p}}$. In order to further make the iterative spectrum estimation procedure robust and stable, the estimated spectrum was expressed as a weighted summation of a set of model spectra $\Omega_{i}(E)$ which were obtained using either Monte Carlo simulation or analytical spectrum generators with different filtration, i.e. the estimated spectrum $\Omega(E)$ can be expressed as follows,
\begin{equation}\label{equ:spek}
\Omega(E)=\sum_{i=1}^{M}c_{i}\Omega_{i}(E).
\end{equation}
with $M$ the number of the model spectra, $c_{i}$ the unknown weights. Based on the model spectra expression, the spectrum estimation problem was formulated as the following iterative optimization problem,
\begin{equation}\label{equ:opt-constraint}
\mathbf{c}=\mathrm{argmin}_{\mathbf{c}}\;\|R_{\mathrm{u}}-\hat{R}_{\mathrm{p}}(\mathbf{c})\|_{2}^{2}, ~~\mathrm{s.t.}~\sum_{1}^{M}c_{i}=1,~\mathrm{and}~c_{i}\geq0.
\end{equation}
The normalization constraint condition $\sum_{1}^{M}c_{i}=1$ and the non-negative constraint condition $c_{i}>0$ were used to normalize the estimated spectrum to unit area and to keep the the solution of~(\ref{equ:opt-constraint}) physically meaningful, respectively.

\subsection{Numerical implementation}


Based on the above subsections, the detailed numerical implementation procedures of the proposed BHC algorithm went as follows:
\begin{enumerate}[label={\arabic*.}]
\item Estimate the effective polychromatic spectrum using either the raw projection data or the uncorrected image volume as input.
\item If the method takes the raw projection data as input, then an initial CT reconstruction using the those raw data was performed to generate beam hardening artifacts contaminated images.
\item Use the OTSU algorithm to segment the uncorrected images into different components and assign attenuation coefficients to these components. These segmented images will serve as template images to yield $\delta_{\mathrm{BHC}}$, an example of how the segmented image looks like is given in Fig.~\ref{fig:reproj}(b).
\item Reproject the template image using a predetermined polychromatic and monochromatic X-ray spectra, the difference of the monochromatic reprojection data and the polychromatic reprojection data is $\delta_{\mathrm{BHC}}$, which is then scaled by $R_{\mathrm{u}}/R_{\mathrm{p}}$ to mitigate segmentation error.
\item If the method is implemented in the projection domain, the scaled mapping term is then added directly to the raw projection data to yield the corrected projection data. The beam hardening artifacts reduced image is reconstructed using the corrected projection data.
\item If the method is implemented in the image domain where the access to the raw projection data is limited, beam hardening artifacts error image is reconstructed from the scaled $\delta_{\mathrm{BHC}}$, after which the error image is added to the uncorrected image to yield the beam hardening corrected image. Note that if the uncorrected image is in HU number, the error image should be calibrated before adding to the uncorrected image.
\end{enumerate}

Note that accurate calculation of $\delta_{\mathrm{BHC}}$ needs an adequate model of the spectrum. When spectrum can not be accurately modeled, it is better to perform BHC in projection domain.
The proposed method can intrinsically correct high order beam hardening artifacts induced by multi-material object. Once a material was segmented from another material, it was assigned attenuation coefficient and then took the polychromatic attenuation process into account, i.e. the difference of monochromatic reprojection and polychromatic reprojection can characterize the nonlinear attenuation of this material. After this, beam hardening artifacts caused by this material can be reduced. This made the method can correct high order beam hardening artifacts straightforward.

In case a material can not be segmented from its background, which meant the material had similar polychromatic attenuation with its background,  we can substitute the material with its background in reprojection. The polychromatic reprojection data was still a good approximate of the real projection, which yielded a correction term that was close to the real value.

\subsection{Simulations}

\begin{figure}
    \centering
    \includegraphics[width=3.0 in]{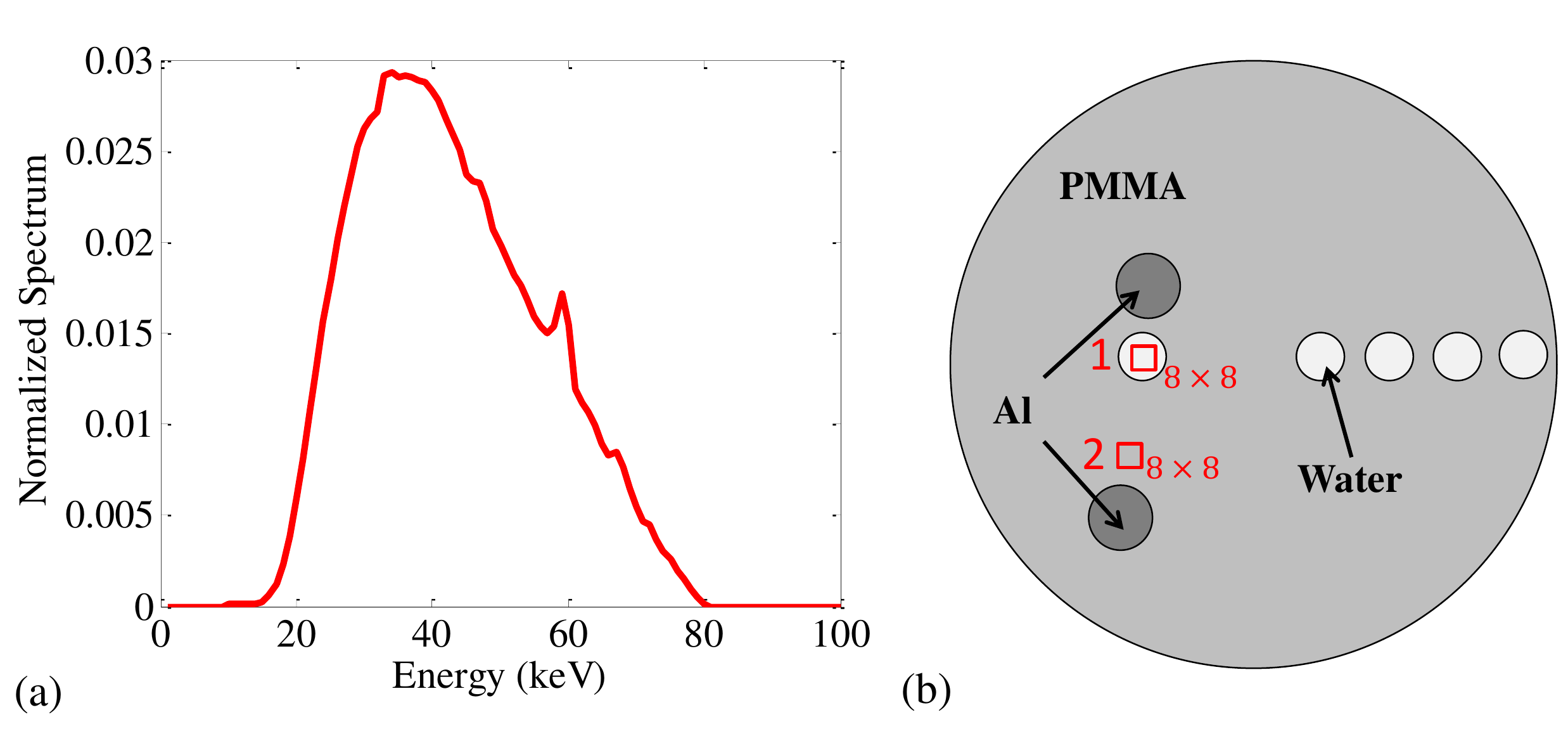}
    \caption{X-ray spectrum and phantom for the numerical simulations. (a) The energy spectrum was simulated using Geant4 with low energy extension package. To yield significant beam hardening artifacts, a small amount of inherent filtration was used in the simulation. (b) The PMMA phantom consists of water and Al inserts. The Al inserts are used to generate high order beam hardening artifacts. The water inserts were added to test the influence of the segmentation on the accuracy of the final correction results, as water and PMMA have the similar attenuation coefficients and can't be differentiated from each other in the segmentation procedure.}
    \label{fig:numericalSim}
    \vspace{-1em}
\end{figure}

To evaluate the proposed algorithm, we performed the numerical simulations using a 2D fan-beam CT geometry. The distance from the source to the center of rotation was 560 mm and the distance from the source to the detector was 740 mm. A circular scan was simulated and a total of 720 projections per rotation
 were acquired in an angular range of $360^{\circ}$. The detector pixel size was 0.254 mm and the detector had 512 pixels.  The X-ray energy spectrum was 80kVp and it was generated using the Geant4 Monte Carlo simulation toolkit with the low energy electromagnetic process extension package~\cite{Allison2006}. Electrons with 80keV constant energy were emitted to hit the tungsten target to yield X-ray photons. The photons were scored after filtrated by 3 mm aluminum and 3 mm oil intrinsic filters. The phantom was a PMMA cylinder with 2 aluminum inserts and 5 water inserts. The diameter of the PMMA cylinder, the water insert and the aluminum insert were 88.29 mm, 6.5 mm and 9.8 mm, respectively. The X-ray energy spectrum and the phantom were shown in Fig.~\ref{fig:numericalSim}. In the simulation, noise was considered and the energy of the monochromatic spectrum used for BHC was set to 39 keV.

\subsubsection{Robustness against attenuation coefficient mismatch}

As mentioned before, we needed to assign attenuation coefficients to each of the segmented components. These values were usually obtained from the NIST data base which were regarded as the standard attenuation coefficients. However, in realistic applications, the real attenuation coefficients may deviate from the standard values. For example, a fatty body may have lower attenuation coefficients than the standard values. Thus, it was necessary to investigate the robustness of the proposed method against the assigned attenuation coefficients. In this evaluation, the raw projection data were generated using the standard attenuation coefficients, while for the correction, all of the assigned attenuation coefficients were scaled to $90\%$ and $110\%$ of the standard values on purpose. The results of BHC with the mismatched attenuation coefficients were compared to the results of BHC with the standard attenuation coefficients.

\subsubsection{Robustness against polychromatic spectrum mismatch}

Since the polychromatic reprojection data was highly related to the polychromatic spectrum, the estimated spectrum $\Omega(E)$ should be accurate enough. However, in routine applications, due to the presence of the scatter radiation or the mismatched attenuation coefficients, $\Omega(E)$ may not be as exact as the true spectrum. Thus it was necessary to investigate the robustness of the proposed method with respect to the estimated polychromatic spectrum. In this evaluation, the BHC method was implemented using mismatched spectra intentionally. The mismatched spectra were simulated using Geant4 with applying more aluminum filtration or less aluminum filtration than the filtration that corresponded to the true spectrum. The results of BHC with the mismatched polychromatic spectra were compared to the results of BHC with the true spectrum.

\subsubsection{Triple-material BHC}
To demonstrate the feasibility of the proposed BHC method for more than two materials, we replaced two of the water inserts with cortical bone inserts. In this case, three-component image segmentation was performed to identify PMMA, cortical bone and aluminum.

\begin{figure}[t]
    \centering
    \includegraphics[width=2.2in]{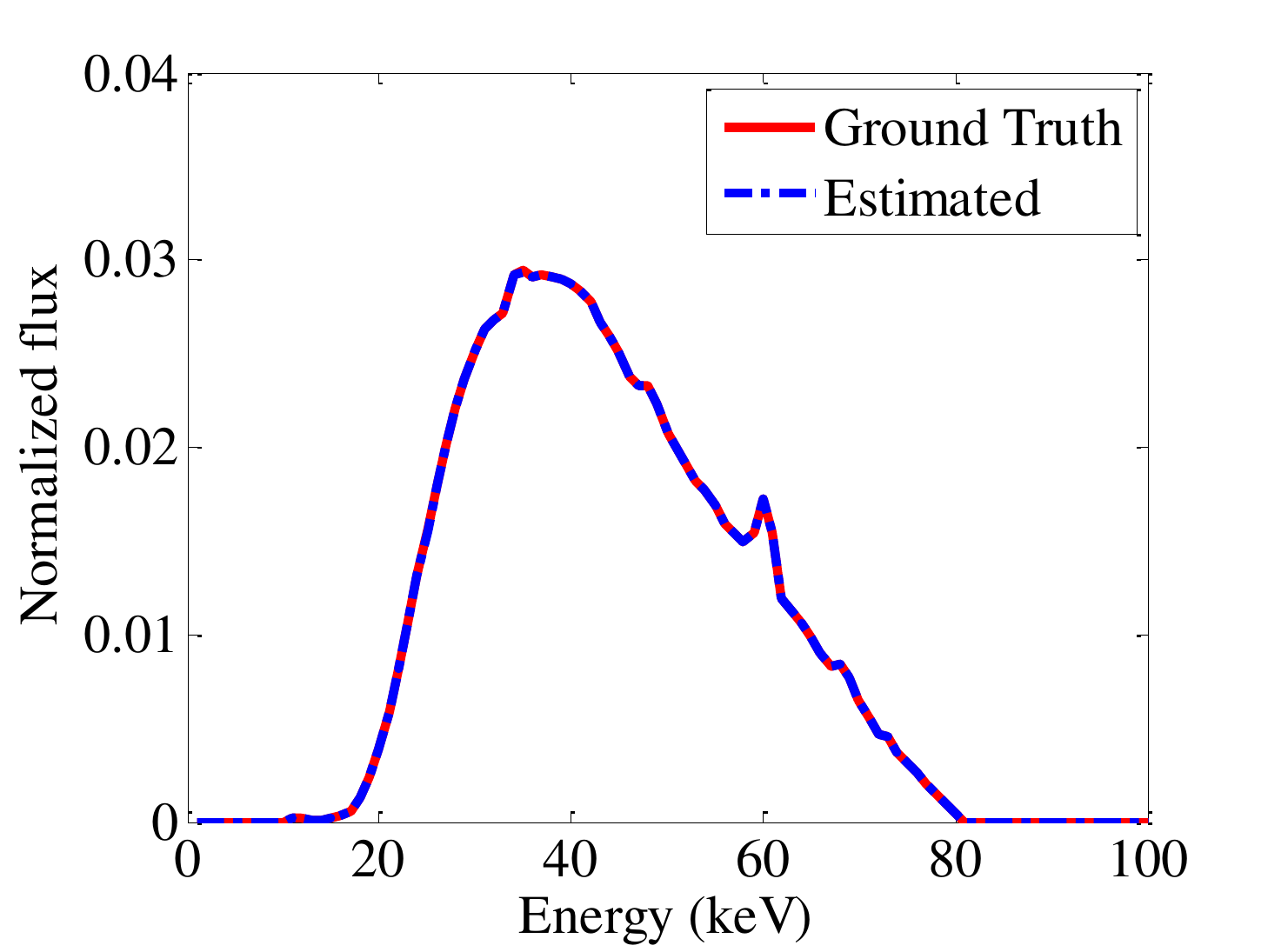}
    \caption{80 kVp X-ray spectrum estimated using the numerical phantom with four model spectra. Since both of the incident X-ray spectrum and the model spectra are generated using MC simulation toolkit Geant4 with the same number of events. The estimated spectrum matches the incident spectrum (ground truth) quite well. }
    \label{fig:spekEst}
    \vspace{-1em}
\end{figure}

\subsubsection{Comparison study}
To show the merit of the proposed method, we had also compared it with one of the state-of-the-art BHC methods, i.e., the empirical beam hardening correction (EBHC) method~\cite{Kyriakou2010}. EBHC is very robust and can be applied in image domain. Meanwhile, it does not require the knowledge of the spectra or the material involved. Based on these reasons, it is widely used in realistic applications.

\subsubsection{Quantitative evaluation}
In order to quantitatively evaluate the performance of the proposed method, we measured HU value and contrast-to-noise ratio (CNR) of a region-of-interest (ROI) within the artifacts. The size of the ROI (shown in Fig.~\ref{fig:numericalSim}(b)) used for measurements was $8\times8$. The CNR of the ROI is defined as:
\begin{equation}\label{equ:cnr}
\mathbf{CNR}=\frac{|\langle \mu_{\mathbf{ROI_1}}\rangle-\langle \mu_{\mathbf{ROI_2}}\rangle|}{\frac{1}{2}(\sigma_{\mathbf{ROI_1}}+\sigma_{\mathbf{ROI_2}})},
\end{equation}
with $\langle \mu_{\mathbf{ROI_i}}\rangle$ and $\sigma_{\mathbf{ROI_i}}$, $~i=1,2$ the mean HU value and the standard deviation of the ROI, respectively.

\begin{figure}[t]
    \centering
    \includegraphics[width=3.0in]{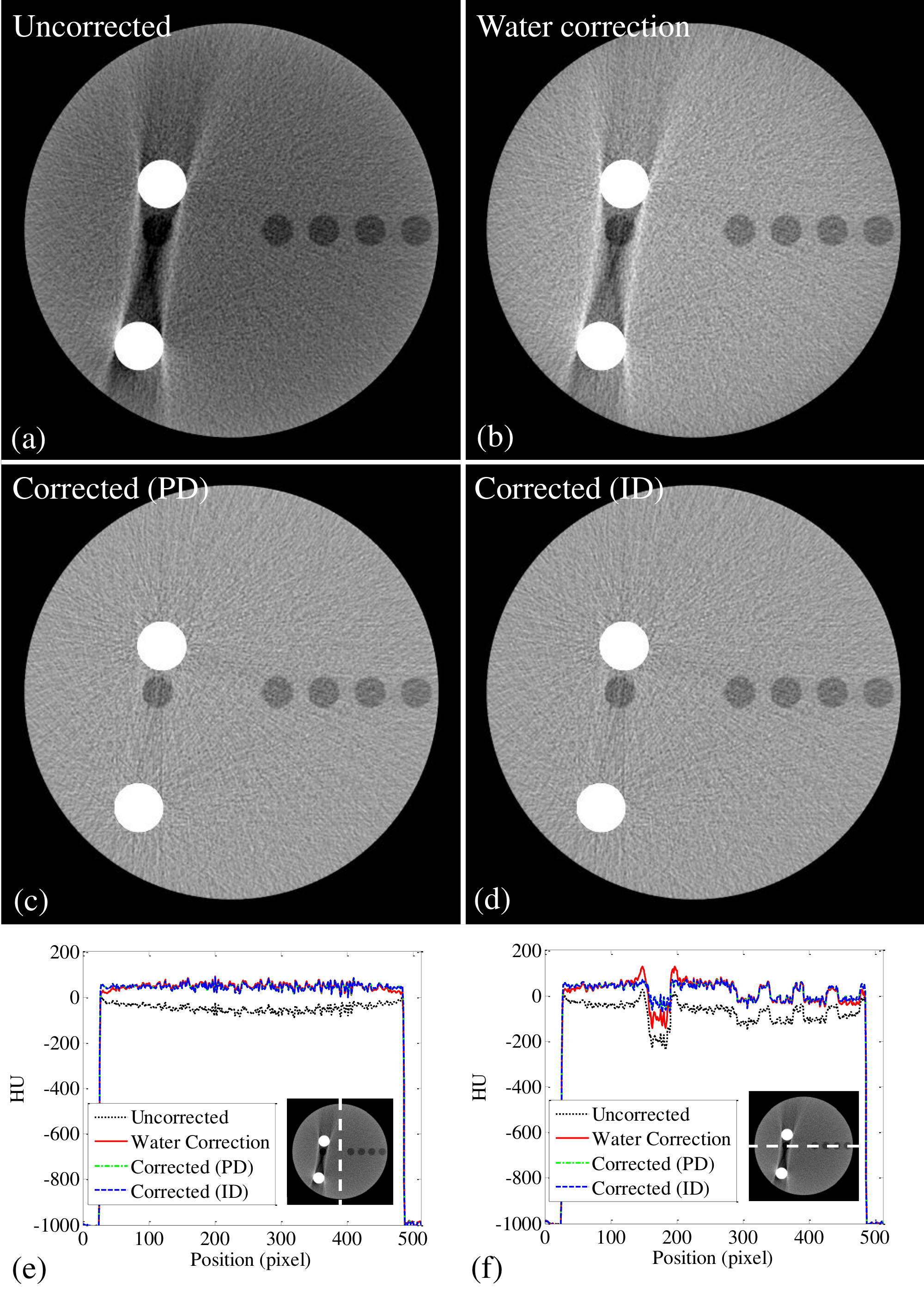}
    \caption{The numerical PMMA phantom with dense inserts and low contrast inserts without (a), with first order correction (b), and with high order correction in both projection domain (c) and image domain (d). Water correction reduced first order beam hardening artifacts such as cupping, but can not reduce high order beam hardening artifacts such as streaks. The proposed method reduced both first order and high order beam hardening artifacts within both projection domain (PD) and image domain (ID). Display window for all of the CT images: [-200, 200] HU. Line profiles along vertical and horizontal direction of the numerical PMMA phantom image with and without beam hardening correction are shown in (e) and (f), respectively. As indicated by the profiles, water correction reduce cupping artifacts but can not reduce the dark streaks between the aluminum inserts. Cupping and streaks are clearly removed by the proposed method within both PD and ID, and the HU are successfully recovered.}
    \label{fig:numericalSimRe}
    \vspace{-1em}
\end{figure}

\subsection{Experiments}

The proposed method was also evaluated using an anthropomorphic Lungman chest phantom data, and \emph{in vivo} canine data acquired with a modern cone-beam helical CT scanner (Discovery CT750 HD, GE Healthcare, WI, USA), and a flat-detector C-Arm CT scanner (Artis Zee, Siemens Healthcare, Forchheim, Germany), respectively. The flat-detector cone beam C-Arm scanner was equipped with a focused anti-scatter grid. Therefore, the contribution of the scatter signal to the raw projection data was neglected. A total of 496 number of views had been acquired in an angular range of $198.1^{\circ}$ for the C-Arm scanner. Measurements were performed at a tube voltage of 120 kVp for the diagnostic CT, and 70 kVp for the C-Arm CBCT. The energy of the monochromatic spectrum were selected to be 63 keV and 42 keV for the physical phantom data and canine data, respectively. Due to the limited access to the raw projection data, the Lungman chest phantom data was corrected in image domain.

\section{RESULTS}

The proposed BHC method was evaluated using numerical simulation to show its ability to reduced high order beam hardening artifacts and to demonstrate its robustness against the assigned attenuation coefficients. In addition, physical phantom and \emph{in vivo} animal data which were acquired with a diagnostic spiral CT scanner and a C-Arm flat-detector CBCT scanner, were also used to demonstrate the data and scanner independence of the method.

\subsection{Simulation study}

\begin{figure}[]
    \centering
    \includegraphics[width=3.0in]{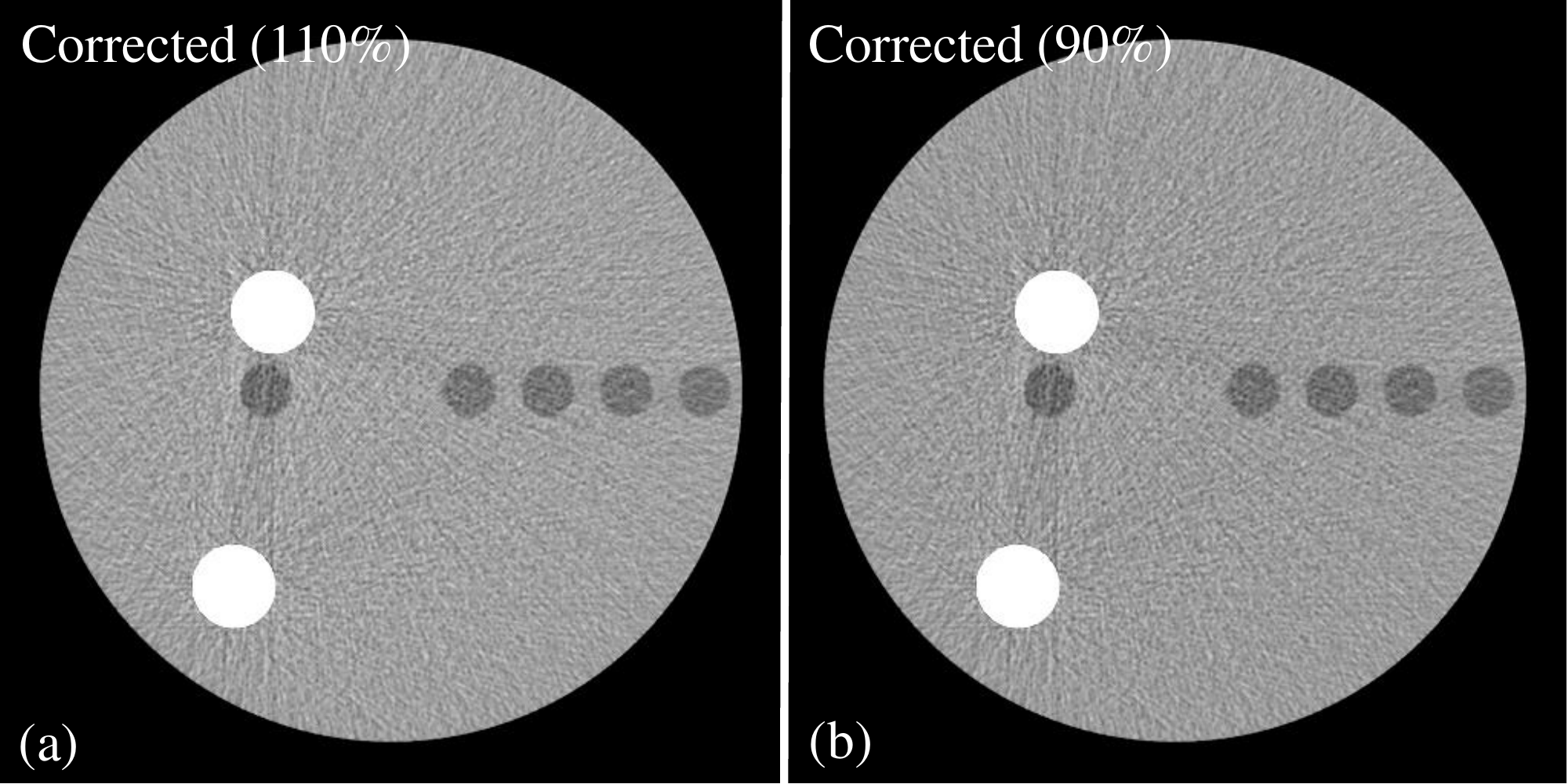}
    \caption{Beam hardening correction of the numerical PMMA phantom using mismatched attenuation coefficients. (a) Correction with $110\%$ of the standard attenuation coefficients for both two segmented components, i.e. aluminum and PMMA. (b) Correction with $90\%$ of the standard attenuation coefficients for both two components. The dark streaks between the aluminum inserts are almost completely removed by the proposed method with mismatched attenuation coefficients, indicating the method is robust with respect to the assigned attenuation coefficients. Display window for the CT images: [-200, 200] HU.}
    \label{fig:robustAtten}
    \vspace{-0.5em}
\end{figure}

\begin{figure}[t]
    \centering
    \includegraphics[width=3.0in]{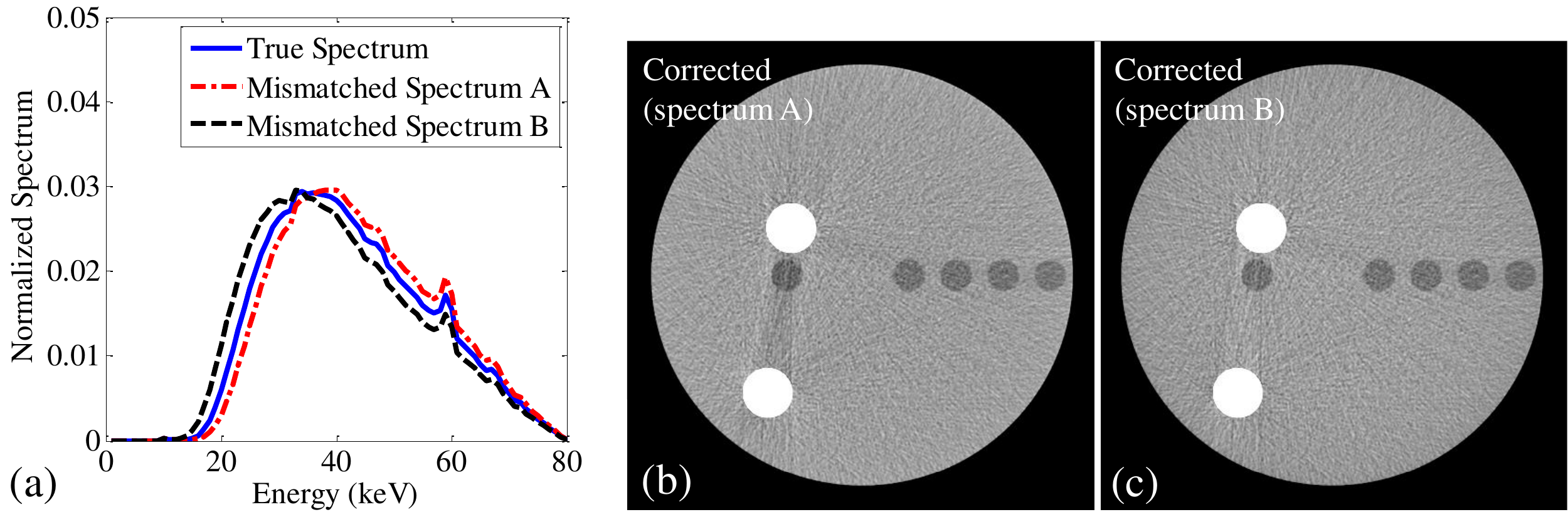}
    \caption{Beam hardening correction of the numerical PMMA phantom using mismatched spectra shown in (a). When BHC was performed using a harder spectrum A which had higher mean energy than the true spectrum, the attenuation of the aluminum inserts was underestimated and residual streaks were presented (b). On the contrary, when a softer spectrum B was used, the attenuation of the aluminum inserts was overestimated and the streaks were over-corrected (c). Display window for the two CT images: [-200, 200] HU.}
    \label{fig:robustSpek}
    \vspace{-1em}
\end{figure}

Figure~\ref{fig:spekEst} shows the 80 kVp polychromatic X-ray spectrum estimated using the uncorrected numerical phantom CT image with 4 model spectra. As can be seen, the estimated spectrum fits the true spectrum quite well. Note that the true incident spectrum was generated using Geant4 with 3 mm aluminum and 3 mm oil filtration, and the model spectra employed in the spectrum estimation were also generated using Geant4 with filtration from 2 mm to 5 mm aluminum. Both of the true spectrum and the model spectra were simulated with the same number of electron events.

Figure~\ref{fig:numericalSimRe} shows the correction results for the cylindrical PMMA numerical phantom with aluminum and water inserts. In the water correction image, the first order beam hardening artifacts (cupping) are reduced, but high order beam hardening artifacts which show as streaks between the aluminum inserts can not be well reduced. This is because water correction treats the aluminum inserts as water and does not fully take the attenuation properties of the aluminum inserts into account. The streaks are significantly reduced using the proposed method within both projection domain and image domain. The missing water insert which locates in the prominent streaks is well recovered and the homogeneity of PMMA and water is greatly restored.


Line profiles along the vertical direction and the horizontal direction are shown in Fig.~\ref{fig:numericalSimRe}~(e) and (f) for the numerical phantom without and with correction. The profiles show that water correction can reduce cupping artifacts, but can not reduce streaks. The proposed method significantly reduces cupping artifacts, as well as streaks, and restores the true HU values, even though the water inserts are not differentiated from the PMMA background in the segmented image, as illustrate in Fig.~\ref{fig:reproj}~(b).

\subsection{Robustness evaluation}

In this section we investigate the influence of mismatched spectrum and attenuation coefficients on the correction quality. The estimate spectrum may deviate from the true spectrum when scatter is present.~
Figure~\ref{fig:robustAtten} shows the correction results of the numerical cylindrical PMMA phantom using mismatched attenuation coefficients. As can be seen, when the assigned attenuation coefficients are $10\%$ larger or smaller than the standard attenuation coefficients, a significant removal of beam hardening artifacts is still achieved, indicating the proposed method is robust against the assigned attenuation coefficients.

Correction results of the numerical phantom using mismatched spectra are depicted in Fig.~\ref{fig:robustSpek}. When a harder spectrum A whose mean energy is higher than the true spectrum is used to perform the polychromatic reprojection, the attenuation of the aluminum inserts is underestimated and streaks between the aluminum inserts are not well compensated. On the contrary, correction with a softer spectrum B would overestimate the attenuation of the aluminum inserts, causing an over-correction result. Note that the corrected CT images are recalibrated using the water correction image to make sure that the PMMA of the corrected image has the same HU as the PMMA of the water correction image.

\subsection{Comparison study}

\begin{figure}[t]
    \centering
    \includegraphics[width=3.0in]{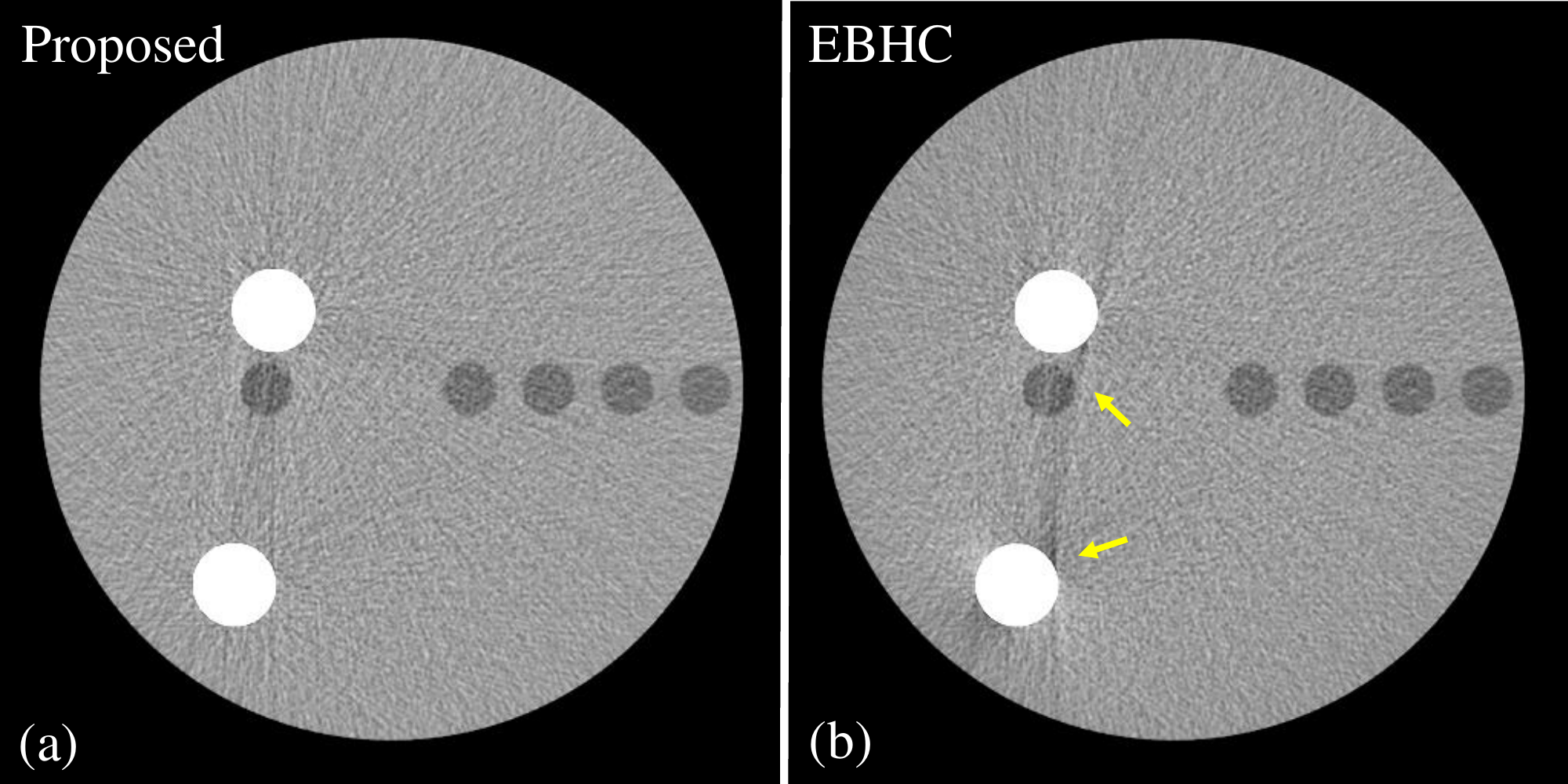}
    \caption{Beam hardening correction of the numerical PMMA phantom using EBHC and the proposed method. (a) Correction using the proposed model-based method. (b) Correction using EBHC. Display window for the CT images: [-200, 200] HU.}
    \label{fig:ebhc}
    \vspace{-0.3em}
\end{figure}

Figure~\ref{fig:ebhc} shows comparison results of correction using EBHC and the proposed model-based method. There are residual streaks between the dense object in the EBHC image and the proposed method outperforms EBHC method. This can be attributed to full utilization of the physics (the X-ray spectrum) while no physics is involved in the EBHC correction. Quantitative evaluation for both methods using HU accuracy and CNR for the ROI in artifacts are shown in Table~\ref{tab:qa}.

\vspace{-0.5em}
\subsection{Triple-material BHC}

\begin{figure}[t]
    \centering
    \includegraphics[width=3.0in]{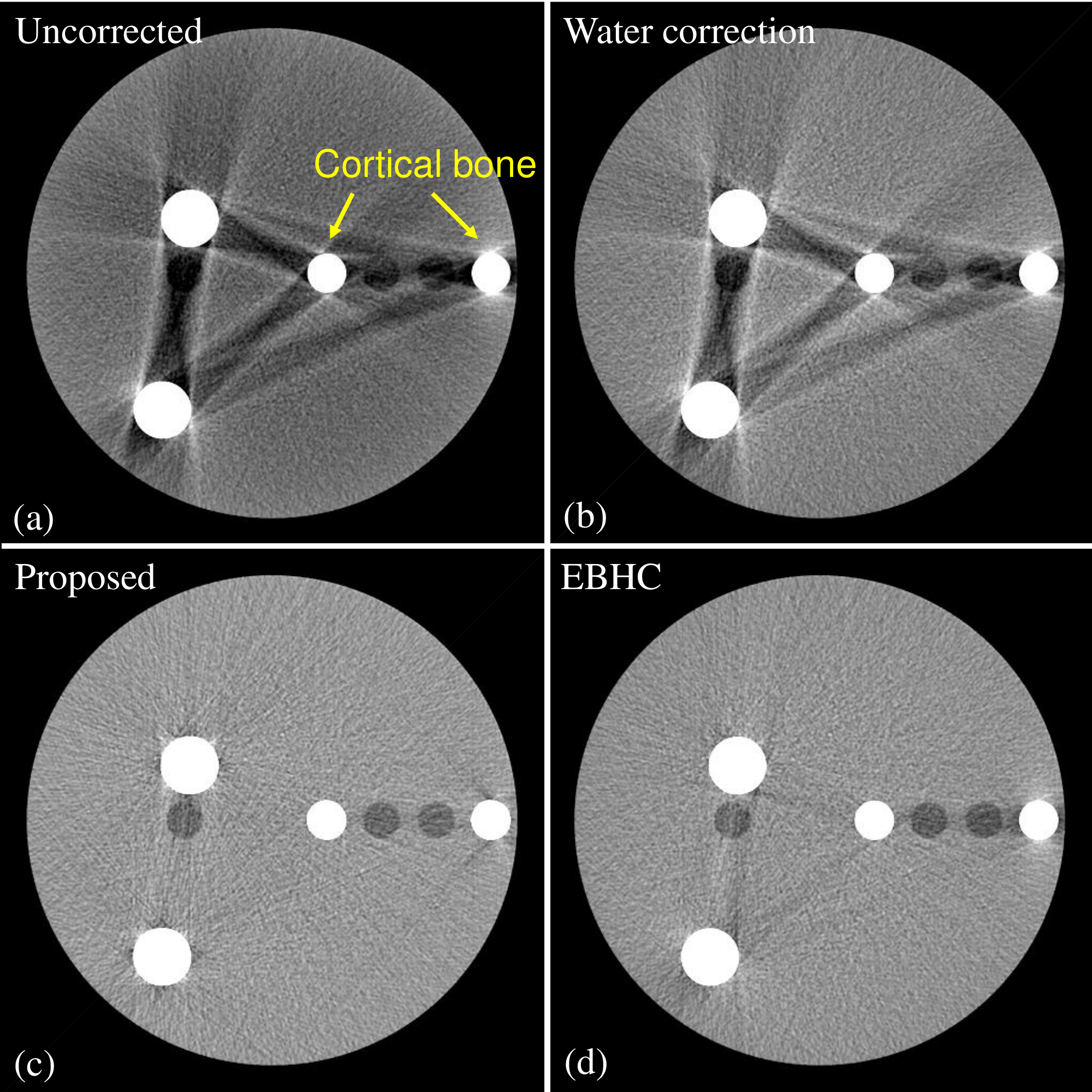}
    \caption{Beam hardening correction of the numerical triple-material phantom using EBHC and the proposed method. (a) Uncorrected image; (b) Image with water correction; (c) Correction using the proposed model-based method; (d) Correction using EBHC. Display window for the CT images: [-200, 200] HU.}
    \label{fig:mmbhc}
    \vspace{-1em}
\end{figure}

BHC for more than two materials is shown in Fig.~\ref{fig:mmbhc}. In this study, we replace two of the water inserts with cortical bone inserts (Fig.~\ref{fig:mmbhc}~(a)), and these two inserts introduce new streak artifacts which can not be reduced by water correction (Fig.~\ref{fig:mmbhc}~(b)). After correction using both the proposed method and EBHC, the streaks are almost completely removed. Quantitative analysis for BHC using different methods are shown in Table~\ref{tab:qa}.

\begin{table}[bp]
\vspace{-1.8em}
\caption{Quantitative analysis of BHC for two-material and three-material using different methods. The ground truth HU value for ROI$_1$ is zero.}
\vspace{-1em}
\label{tab:qa}
\begin{center}
\begin{tabular}{|p{15mm}|c|c|c|c|c|}
\hline
\multicolumn{2}{|c|}{\multirow{2}{*}{Correction methods}} & \multicolumn{2}{ c| }{Two-material} &\multicolumn{2}{ c| }{Triple-material} \\ \cline{3-6}
\multicolumn{2}{|c|}{}
 & HU (ROI$_1$) & CNR & HU (ROI$_1$) & CNR \\
\hline
\multicolumn{2}{ |c| }{Uncorrected} & -167 & 2.08 & -168 & 2.71 \\
\hline
\multicolumn{2}{ |c| }{Water correction} & -119 & 2.2 & -120 & 2.78 \\
\hline
\multicolumn{2}{ |c| }{Corrected (PD)} & 0 & 3.94 & 0 & 3.81 \\
\hline
\multicolumn{2}{ |c| }{Corrected (ID)} & 0 & 3.94 & 0 & 3.81 \\
\hline
\multicolumn{2}{ |c| }{EBHC} & 0 & 3.0 & 0 & 3.57 \\ \hline
 Mismatched  & 110\%  & 3 & 3.94 & 3 & 3.81 \\ \cline{2-6}
           materials   & 90\%  & -4 & 3.93 & -4 & 3.80 \\ \hline
 Mismatched  & A  & -18 & 3.77 & -4 & 3.73 \\ \cline{2-6}
           spectra   & B & 12 & 4.12 & 5 & 3.91 \\ \hline
\end{tabular}
\end{center}
\end{table}

\vspace{-1em}
\subsection{Experimental studies}

\begin{figure*}[t]
    \centering
    \includegraphics[width=0.8\textwidth]{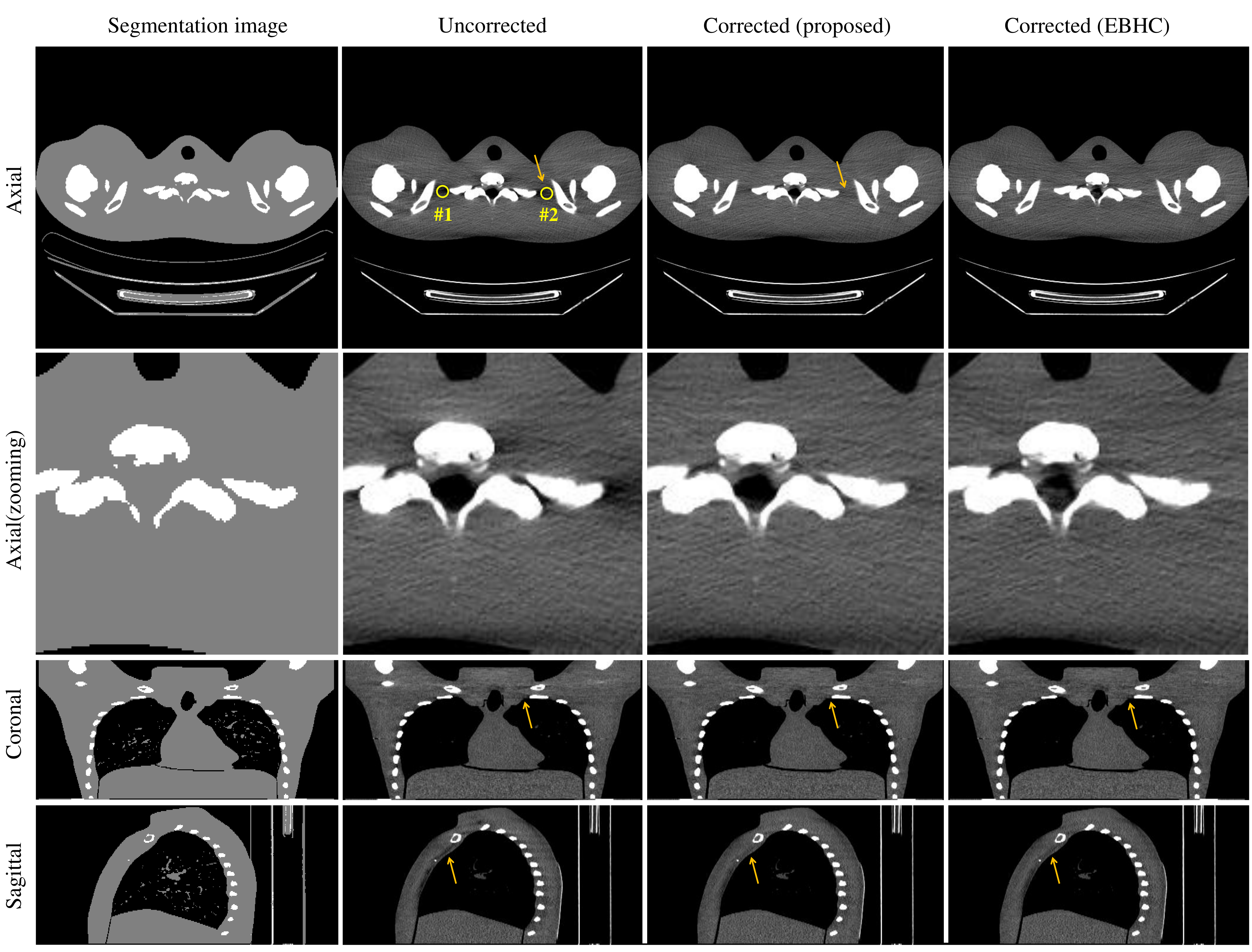}
    \caption{Lungman chest phantom images without and with correction for Discovery CT750 HD helical CT. The dark streaks between the bones and ribs at the shoulder are almost completely reduced after correction. The labelled regions of interest (\#1, \#2) were employed to evaluate of HU accuracy. Display window for the CT images: [-100, 200] HU.}
    \label{fig:lungmanPhantom}
    \vspace{-1em}
\end{figure*}

\begin{figure*}[t]
    \centering
    \includegraphics[width=0.8\textwidth]{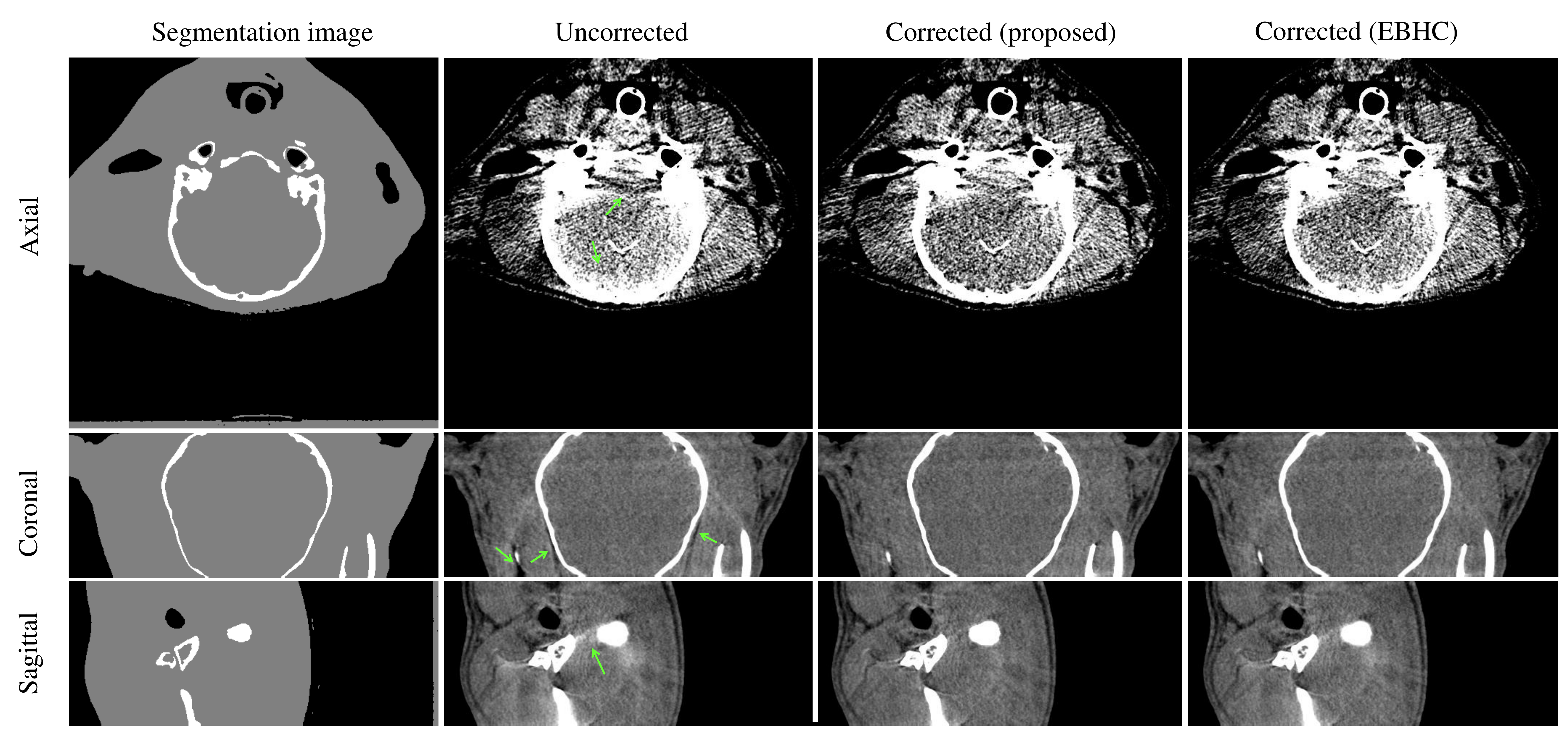}
    \caption{Canine head images without and with correction for C-Arm flat detector CT. The dark streaks between the bones at the skull base which are marked with arrows in the uncorrected images, are significantly reduced after correction. Display window for the axial CT images: [-20, 80] HU. Display window for the coronal and sagittal CT images: [-200, 400] HU. }
    \label{fig:canine}
    \vspace{-1em}
\end{figure*}


Figure~\ref{fig:lungmanPhantom} shows the correction results for the Lungman chest phantom which was scanned using the Discovery CT750 HD helical CT scanner. The segmentation images used for correction are shown in the first column. The prominent dark streaks between the bones and ribs at the shoulder are almost completely removed after correction using both the proposed method and EBHC, and the homogeneity of the phantom materials is restored. The HU values are restored in ROI$_1$ and ROI$_2$ from -55 HU and -48 HU to -9 HU and -7 HU (the proposed method), and to 11 HU and 13 HU (EBHC), respectively. The coronal images suggest the proposed method provides more uniform correction.

Figure~\ref{fig:canine} shows the canine head images acquired by the C-Arm CBCT scanner without and with correction. As can be seen in the axial images which show in a tight window, both the proposed method and EBHC mitigate the dark streaks between the boney structures at the skull base and improve image quality. The proposed method outperforms EBHC with respect to artifacts reduction between the dense boney structures and periphery of the skull (the green arrow in the axial images). In the coronal and sagittal images, the dark streaks are almost completely removed. Beam hardening correction of the different data acquired on different scanners demonstrate the generality of the proposed method. 

%
%
%

\section{DISCUSSION}

We developed a model-based BHC method that accurately corrected high order beam hardening artifacts, such as streaks. The method is based on the modeling of the physical interactions between the X-ray photon and the materials. From this sense, the estimation of an effective polychromatic spectrum which models the nonlinear attenuation of the object is necessary. In this study, we have used the indirect transmission measurement-based spectrum estimation method~\cite{zhao2015}. Note that other spectrum estimation methods can also be employed. For example, the spectrum can be measured using an energy-resolved photon counting detector~\cite{duisterwinkel2015} or estimated using a newly proposed dual-energy CT-based method~\cite{zhao2017}.The spectrum estimation can take either the raw projection data or uncorrected images as input, namely, it can be performed in both projection domain and image domain. The estimated polychromatic spectrum is employed to reproject a segmented volume which is generated from the uncorrected volume. By calculating the difference between the monochromatic reprojection data and the polychromatic reprojection data, one can yield $\delta_{\mathrm{BHC}}$ which has incorporated the nonlinear attenuation of the projection. If the BHC method is implemented in raw projection domain, $\delta_{\mathrm{BHC}}$ is then added directly to the raw projection data to yield the corrected projection data. If the BHC method is implemented in image domain where the access to the raw projection data is limited, then a beam hardening artifacts error volume is reconstructed using $\delta_{\mathrm{BHC}}$. By adding the error volume to the uncorrected volume, we can obtain the corrected volume. To mitigate the impact of segmentation error which increases the inconsistency between $R_u$ and $R_p$, one can scale $\delta_{\mathrm{BHC}}$ using a factor $R_u/R_p$ to partly compensated the missing information induced by inaccurate segmentation.

BHC methods involving material segmentation and forward projection have considered to be effective solutions for bone induced artifacts correction~\cite{joseph1978,Hsieh2000}. These methods perform segmentation first, followed by the forward projection to compensate the contribution of bone. They usually need a lot of calibration measurements to determine the parameters for bone, and if a third material presents, additional calibration measurements are required. They also need to be implemented iteratively to fully address the artifacts, which is computational expensive. By accurately modeling the polychromatic nonlinear attenuation, the proposed method does not involve iterative forward and backward projection. Thus it is desirable for fast implementation. Note that BHC in image domain has been originally proposed in Refs.~\cite{moore1980,crawford1986,crawford1995}. While these works focus on the numerical implementation of reprojection, the proposed work focuses on the incorporation of the physical nature into the correction procedure. In this study, when BHC is performed in image domain, we first forward projection the image to obtain projection data, which is then used for spectrum estimation. If the image is preprocessed using water correction, the estimated spectrum should incorporate the effect of water correction. This might make the image domain BHC complicated. In this case, spectrum estimation using phantoms (such as Catphan phantom) may be helpful.

The computational demand of the algorithm is dominated by the additional reprojection and backprojection. The reprojection is considered having a similar computation time as the backprojection. By employing the modern hardware acceleration techniques, such as GPU, the additional computational load should not be a critical issue. In this study, it took about 100 s to correct the canine data set ($512\times512\times200$) using a NVIDIA GeForce GTX 480 card. Compared to many iterative high order beam hardening techniques which need repetitive forward projections and backprojections, the proposed algorithm is desirable for real-time clinical applications.

The presented algorithm is robust with respect to the assigned attenuation coefficients. When the realistic attenuation coefficients had $10\%$ deviation from the standard NIST values which might be rare in practice, the algorithm can still yield acceptable results. When the algorithm is applied to the data set that was acquired on CBCT, scatter radiation may be a concern which causes an inaccurate effective spectrum. In this study, the experimental data acquired on multi-slices helical and C-Arm CBCT scanners arguably contains a comparably high scatter fraction. However, this did not show considerable influence on the corrected images. This might be attribute to the focused scatter grid employed by the systems, rejecting a considerable amount of scattered radiation. 

As demonstrated in the numerical simulations, the proposed BHC method is robust against segmentation. For the numerical phantom, due to the presence of noise and artifacts, only aluminum and PMMA were identified (shown in Fig.~\ref{fig:reproj}). However, cupping and streaks were still completely removed. This is because the water inserts have similar attenuation properties as the PMMA background. Nevertheless, just as other first order beam hardening correction algorithms which treat tissue as water to perform water correction, this will not be a critical issue for realistic application. For high order beam hardening artifacts induced by the aluminum inserts, the nonlinear attenuation (root cause of streaks) of these inserts are easy to calculate, as the aluminum inserts (dense material) are easy to be identified from the background material.

\section{Conclusion}

In this study, based on spectrum estimation, an efficient beam hardening artifacts reduction method was developed. The method is potentially clinical useful and can be implemented within both projection domain and image domain when there is an adequate model of the spectrum. Numerical simulations, experimental phantom data and \emph{in vivo} animal data which were acquired on different CT scanners, were used to evaluate the method. The results show the proposed method can significantly reduce both first order and high order beam hardening artifacts. Further tests show the method is robust with respect to the attenuation coefficient and it is independent to the data and CT system. This work is promising to be applied to commercial products.



%

%


\section*{Acknowledgment}
The authors are grateful to Drs. Michael Speidel, Ke Li, Sebastian Schafer and Kevin Royalty for assistant of the Lungman chest phantom and the canine data acquisition. The authors also thank Drs. Jie Tang and Xuexiang Cao for their useful discussions.

%

\ifCLASSOPTIONcaptionsoff
  \newpage
\fi



\bibliographystyle{IEEEtran}

%

\end{document}